# Directional Carrier Transport in Micrometer-Thick Gallium Oxide Films for High-Performance Deep-Ultraviolet Photodetection

Wenrui Zhang,[1,2]* Wei Wang,[1] Jinfu Zhang,[1] Tan Zhang,[1] Li Chen,[1] Liu Wang,[1] Yu Zhang,[3] Yanwei Cao,[1] Li Ji,[3] Jichun Ye[1,2]*

[1] Zhejiang Provincial Engineering Research Center of Energy Optoelectronic Materials and Devices, Ningbo Institute of Materials Technology and Engineering, Chinese Academy of Sciences, Ningbo 315201, China
[2] Yongjiang Laboratory, Ningbo 315201, China
[3] State Key Laboratory of ASIC and System, School of Microelectronics, Fudan University, Shanghai 200433, China

## Abstract

Incorporating emerging ultrawide bandgap semiconductors with a metal-semiconductor-metal (MSM) architecture is highly desired for deep-ultraviolet (DUV) photodetection. However, synthesis-induced defects in semiconductors complicate the rational design of MSM DUV photodetectors due to their dual role as carrier donors and trap centers, leading to a commonly observed trade-off between responsivity and response time. Here, we demonstrate a simultaneous improvement of these two parameters in $\varepsilon$-Ga$_2$O$_3$ MSM photodetectors by establishing a low-defect diffusion barrier for directional carrier transport. Specifically, using a micrometer thickness far exceeding its effective light absorption depth, the $\varepsilon$-Ga$_2$O$_3$ MSM photodetector achieves over 18-fold enhancement of responsivity and simultaneous reduction of the response time, which exhibits a state-of-the-art photo-to-dark current ratio near $10^8$, a superior responsivity of >1300 A/W, an ultrahigh detectivity of >$10^{16}$ Jones and a decay time of 123 ms. Combined depth-profile spectroscopic and microscopic analysis reveals the existence of a broad defective region near the lattice-mismatched interface followed by a more defect-free dark region, while the latter one serves as a diffusion barrier to assist frontward carrier transport for substantially enhancing the photodetector performance. This work reveals the critical role of the semiconductor defect profile in tuning carrier transport for fabricating high-performance MSM DUV photodetectors.

## 1. INTRODUCTION

Ultrawide-bandgap (UWBG) semiconductors exhibit competing characteristics for developing next-generation power electronics, ultraviolet photodetectors and sensors.[1-3] As an emerging UWBG material, gallium oxide ($Ga_2O_3$) possesses several appealing advantages for fabricating deep-ultraviolet (DUV) photodetectors, including an inherent wide bandgap ($\geqslant 4.6$ eV) with strong band-edge absorption, a simple binary composition, and a relatively long minority carrier diffusion length.[4,5] As such, extensive research efforts are devoted to fabricating DUV photodetectors based on different forms of $Ga_2O_3$, such as nanostructured $Ga_2O_3$,[6] amorphous $Ga_2O_3$[7-9] , single-crystalline β-,[10-12] ε-,[13-15] α-$Ga_2O_3$ thin films.[16-18] and mixed-phase $Ga_2O_3$.[19-21] A majority of the above-mentioned DUV photodetectors adopt a planar metal-semiconductor-metal (MSM) device structure, as shown in Scheme 1a, owing to the simple device fabrication and feasible integration with external circuits.[22-24] In a typical photodetection incident, photon collection and photocarrier transport are two essential steps (Scheme 1b), which are associated with the effective light absorption length ($d_p$) and the minority carrier diffusion length ($d_h$, hole in the $Ga_2O_3$ case), respectively. Accordingly, these two fundamental parameters provide key design guidelines for optimizing the structure geometry of MSM devices, including the film thickness ($t_{film}$) and the electrode spacing ($t_{space}$), in an effort to obtain maximum photodetector performance.

The principle for optimizing the MSM device geometry is to maximize photon collection and facilitate carrier collection. Compared to the straightforward role of $t_{space}$ that mainly determines the transport distance, designing an appropriate $t_{film}$ is relatively complicated as it relies on the trade-off between light absorption and carrier transport. In fact, very large inconsistency of the device parameters is seen in previous $Ga_2O_3$ MSM photodetector studies, where the reported optimized $t_{film}$ values vary from 30 to 1000 nm.[7-18] While such inconsistency is likely attributed to different material qualities in these studies, it is not clear how the material defect affects the DUV photodetection behavior of a $Ga_2O_3$-based MSM photodetector, and how to coordinate the device's parameters for optimized performance. Particularly, a trade-off effect is commonly observed that relies on lattice defects as trap centers to enhance the photocurrent gain, which, however, inevitably compromises the response speed due to the slow de-trapping process.[25-27] In this work, we perform a systematic study to understand how to design the device's geometric parameters with $Ga_2O_3$ characteristics and defects for fabricating high-performance MSM DUV photodetectors. It is discovered that adopting a micrometer thickness rather than a commonly suggested $d_p$ simultaneously achieves more than 18-fold enhancement of responsivity and over 100% enhancement of the response speed in ε-$Ga_2O_3$ thin film MSM photodetectors. The simultaneous optimization of the responsivity and the response time is achieved via establishing an extra dark region in the micrometer-thick films as a diffusion barrier, which drives directional carrier transport to the front electrode for larger carrier collection efficiency in a non-trapped fast manner.

## 2. RESULTS AND DISCUSSION

### 2.1 Thin film growth and structure characterization

The 1%Sn-doped $Ga_2O_3$ (molar ratio, Sn-$Ga_2O_3$) films are grown using pulsed laser deposition (PLD) on c-plane single-crystal sapphire substrates. The Sn doping is critical to assist the formation of ε-$Ga_2O_3$ films rather than contributing excessive carriers in current synthesis conditions.[28] A range of laser pulse numbers from 600 (0.6kp) to 40000 (40kp) pulses is used to fabricate $Ga_2O_3$ films with various film thicknesses. The X-ray diffraction (XRD) results in Figure 1a show that all the films are crystallized as ε-$Ga_2O_3$ with a single set of (00$l$) diffraction peaks in the out-of-plane direction. The film crystallinity increases at thicker films, as seen from the rocking curve measurements in Figure 1b. The full width at half maximum (FWHM) of the (004) ε-$Ga_2O_3$ peak decreases from 1.15 degree to 0.62 degree as the laser pulse number increases from 0.6kp to 40kp. The film thickness increases almost linearly with the pulse number (Figure S1), which is established by checking the cross-sectional specimens using the scanning electron microscopy (SEM). Figure 1c presents two representative SEM images for measuring the film thickness for 2kp and 20kp, which determines an average growth rate of 0.10-0.12 nm/pulse. The laser pulse number of as-deposited ε-Ga$_2$O$_3$ films is used in the following discussion to accurately label the film thickness.

The film microstructure is further characterized by transmission electron microscope (TEM) and energy-dispersive X-ray (EDX) mapping. It is observed uniform single-phase structure of the ε-Ga$_2$O$_3$ film from the cross-sectional scanning transmission electron microscopy (STEM) image and corresponding EDX maps in Figure 1d. The high-resolution TEM image in Figure 1e reveals a homogenous crystalline film lattice accompanied with a disordered interlayer region near the interface. Such interlayer region could help accommodate the large lattice mismatch between the film and the substrate, which is widely observed in previous reports.[29,30] In addition to the decent phase purity and crystallinity, these ε-Ga$_2$O$_3$ films are quite smooth, exhibiting surface roughness (R$_a$) values of 0.49 nm (1.61 nm) for the 2kp (20kp) film (Figure 1f).

## 2.2 Thickness-dependent photodetector performance

Optical absorption measurements reveal sharp absorption edges and determine a direct optical bandgap of 4.9 ± 0.1 eV for ε-Ga$_2$O$_3$ (Figure 2a and Figure S2). Analysis of the absorption data of the 0.6kp film determines the absorption coefficients larger than $1.2 \times 10^5$ cm$^{-1}$ for the light with a wavelength λ < 260 nm, which is consistent with previous reports.[31] This identifies a $d_p$ value of ~200 nm that is close to the 2kp thickness, assuming more than 90% absorption of the incident light (Figure 2b). Since further thicker part may be treated as a dark region, the conventional design strategy would expect a maximum photocurrent occurring near the 2kp thickness of the ε-Ga$_2$O$_3$ film. To testify this strategy, we prepare MSM photodetectors by depositing metallic Au/Ti electrodes on top of ε-Ga$_2$O$_3$ films with a specific finger spacing (500 μm), length (150 μm) and width (150 μm). Surprisingly, it is observed that the photocurrent increases significantly even when $t_{film}$ exceeds the 2kp value, while the dark current remains at an extremely low level below 10$^{-12}$ A for all investigated film thicknesses. Under 254 nm illumination with an intensity of 4 mW/cm$^2$ and a bias of 20 V, a photocurrent over $3.0 \times 10^{-5}$ A is obtained for ε-Ga$_2$O$_3$ photodetectors with $t_{film}$ of 15kp and thicker ones, representing more than 18-fold enhancement than the 2kp film (Figure 2c). Similar thickness dependence of photocurrent is also seen with a much weaker light intensity of 20 μW/cm$^2$ (Figure 2d and Figure S3), which demonstrates the reproductivity of this intriguing phenomenon under both strong and weak light conditions.

To fully understand the performance difference affected by $t_{film}$, we select ε-Ga$_2$O$_3$ photodetectors with two representative thicknesses of 2kp and 20kp for more comprehensive photoelectric characterization. Figure 3a and 3b present the current-voltage (*I-V*) curves of these two photodetectors under 254 nm illumination with various intensities from weak (4 μW/cm$^2$) to strong (9.3 mW/cm$^2$). Both photodetectors present clear photoresponse for the investigated range of light intensity. A noticeable ON-OFF ratio of ~1000 remains for the 20kp photodetector even at a weak light intensity of 4 μW/cm$^2$, implying decent capability for weak light detection. These two photodetectors both follow a linear power dependence with a R$^2$ > 0.99 for the entire intensity range (Figure 3c), exhibiting a broad linear dynamic range highly desired for practical light detection.[32] The dynamic photoresponse of both photodetectors is relatively fast (Figure 3d and 3e). Based on the transient photoresponse to 254 nm illumination of 4 mW/cm$^2$, the 20kp photodetector delivers a rise time ($t_{rise}$) of 0.166 s and a fall time ($t_{fall}$) of 0.075 s, compared to 0.342s ($t_{rise}$)/0.061s ($t_{fall}$) for the 2kp-thick photodetector. It is somewhat surprising that the thicker film achieves larger responsivity and an even faster rise time, compared to a more-commonly observed trade-off relation of these two metrics. The wavelength-dependent photoresponse of the ε-Ga$_2$O$_3$ photodetectors in Figure 3f exhibits good solar-blind photodetection behavior. A peak responsivity ($R_p$) is obtained near the wavelength of ~240 nm, and the 20kp photodetector gives a $R_p$/R(400nm) rejection ratio of $1.87 \times 10^4$.

### 2.3 Electrode spacing effect and photodetector optimization

Besides the film thickness $t_{film}$, the top electrode spacing ($t_{space}$) is another key design parameter that determines the carrier transport distance. Adjusting $t_{space}$ towards the hole diffusion length $d_h$ of Ga$_2$O$_3$ is helpful to shorten the transit distance and facilitate carrier collection. To optimize this parameter, we select the 20kp-thick ε-Ga$_2$O$_3$ film and fabricate MSM photodetectors with $t_{space}$ varying from 1 μm to 200 μm, as shown in Figure 4a. The spacing of 1 μm is near the minimum feature size of the conventional lithography process, and 200 μm is among the largest electrode sizes used in most Ga$_2$O$_3$-based MSM photodetectors. The dark current remains relatively stable less than $5 \times 10^{-12}$ A for all the spacings, reflecting ultralow point defect level for these ε-Ga$_2$O$_3$ films (Figure S4). By contrast, the photocurrent exhibits more apparent dependence on $t_{space}$ (Figure 4b). The photodetector with $t_{space} = 1$ μm exhibits a photocurrent of ~$1.1 \times 10^{-4}$ A, which is nearly 10 times larger than $1.3 \times 10^{-5}$ A observed for $t_{space} = 150$ μm. This gives a state-of-the-art photo-to-dark current ratio (PDCR) of $9.49 \times 10^7$. Incorporating the effective electrode area for the responsivity (detectivity) calculation, the ε-Ga$_2$O$_3$ photodetector with $t_{space} = 1$ μm achieves a maximum responsivity (detectivity) of 1388 A/W ($1.01 \times 10^{16}$ Jones), which is the highest among previous reported MSM DUV photodetectors (Table S1).[7-18] The response time is also affected by $t_{space}$, with mucher longer $t_{rise}$ observed for samller $t_{space}$ (Figure 4c). Ovearll, as $t_{space}$ increases, the responsivity and detectivity decrease continuously (Figure 4d), which is accompanied by an obvious increase with the response speed, in particular for the rise time (Figure 4e). Such trade-off relation tuned by $t_{space}$ is distinct from the relation by $t_{film}$, reflecting a fundamentally different role of the film thickness in assisting carrier transport.

## 2.4 Defect profile characterization

Since carrier transport is closely related to the defect density and distribution, we probe such information using depth profile X-ray photoelectron spectroscopy (XPS) to understand the key impact of the extra film thickness in heteroepitaxial ε-$Ga_2O_3$ films on simultaneously enhancing the responsivity and the response speed. Using the 2kp film as a representative sample, we collect XPS spectra at regions from the top surface to the bottom substrate. The C *1s* peak from the surface absorbed carbon and the Al 2*p* peak from the substrate are representative top and bottom signals for monitoring the etching process (Figure 5a). As a major defect indicator, the oxygen vacancy is widely analyzed from the deconvolution of the O 1*s* peak, which consists of a peak signal from the lattice oxygen near 530 eV and the oxygen vacancy signal near 532 eV.[33,34] We start discussing the spectra evolution from an etching time of 300 s to avoid the complication of surface-absorbed hydroxides. For the 2kp film, it is observed clear peak broadening or appearance of an apparent peak shoulder near 532 eV with increasing etching time (Figure 5c), compared to much less variation of Ga 2*p* or Sn 3*d* peaks (Figure S5). Based on the etching distance, the width of this defective region is found larger than 100 nm starting from the substrate. Once the etched position arrives at the substrate, the O 1*s* peak becomes narrow again as the main contribution comes from the lattice oxygen in $Al_2O_3$. We further perform the depth profile XPS analysis for the 20kp film using the same etching depth as 2kp, so that it mainly probes its top region (Figure 5b). The O 1s peak signal is found mainly from the lattice oxygen (Figure 5d), reflecting a uniform and more defect-free top region. Stronger XPS signals for the Sn elements at the surface is seen from for both 2kp and 20kp films (Figure S5d), which again suggests a catalyst role of the Sn element in assisting the growth of the ε-Ga$_2$O$_3$.[35,36]

The distinct defect distribution between the interface region and the top film provides valuable insights for understanding the dramatical enhancement of responsivity with faster response time occurring at a thickness far larger than $d_p$. It is well established that material defects, in the form of lattice disorders and surface defects, widely exist for films grown on heterogeneous substrates.[37,38] This is particularly true for the film growth with large lattice mismatch, such as the case of ε-Ga$_2$O$_3$ on sapphire in this study. The STEM image in Figure 6a identifies a disordered interface region with a high density of misfit dislocations and domain boundaries, in particular for the initial 100-150 nm region, which is consistent with the XPS observation. Such defects do not form a conductive channel through the film, so that a similarly low level of dark current is observed for films with different thicknesses. In the thicker film region, many dislocations get annihilated or terminated which leads to larger domain size and higher film crystallinity (Figure S6). In a thin film case with $t_{film} \lesssim d_p$, the defective interface region largely overlaps with the light absorption region. Thus, a considerable amount of photocarriers are more easily trapped and recombined with free carriers in this region, which seriously compromises the device photocurrent (Figure 6b). By contrast, for thicker films ($t_{film} > d_p$), an extra dark region with higher crystallinity and less defect density is established, as observed in the case of the 20kp film. Such a dark region could hinder the backflow of photocarriers and drive them to diffuse towards the front contact, thus enhancing the carrier collection efficiency and the overall photocurrent (Figure 6c). Such diffusion-driven photocarrier transport is seen in semiconductors with large exciton binding energy and long minority carrier diffusion lengths, such as Ga$_2$O$_3$, ZnO, GaN and AlGaN.[39-43]

A key advantage of this transport behavior is that the extra photocurrent gain presents less dependence on the formation of trap centers, and the directional carrier diffusion within the low-defect region could allow simultaneous optimization of the response speed. We note that this transport behavior is different from trap-assisted transport due to lattice disorders or surface defects that otherwise cause a trade-off between the responsivity and the response time. The latter phenomenon is actually seen in the optimization trend of $t_{space}$ in this study, where surface defects likely play a more important role in dictating the photodetector performance at a smaller feature size. Figure 6d summarizes the key device metrics of our $\varepsilon$-$Ga_2O_3$ photodetector compared with previously reported state-of-the-art $Ga_2O_3$ MSM photodetectors, evidencing its superior performance in simultaneous achievement of a record high responsivity and PDCR, a low dark current, microsecond decay time and a decent UV-visible rejection ratio.

## 3. CONCLUSIONS

We have developed high-performance MSM DUV photodetectors based on heteroepitaxial $\varepsilon$-$Ga_2O_3$ films by establishing an extra dark region with a low defect density to significantly enhance carrier collection efficiency. The optimized micrometer-thick $\varepsilon$-$Ga_2O_3$ MSM photodetector achieves a record high PDCR of $9.48 \times 10^7$ and a responsivity of 1388 A/W, a relatively fast decay time of 123 ms, a pA dark current under a 20 V bias, and a UV-Vis rejection ratio of $1.87 \times 10^4$. Combined TEM and depth profile XPS investigations reveal a broad defective region stemming from the $\varepsilon$-$Ga_2O_3$/$Al_2O_3$ interface, in neighboring with a more defect-free region in the upper portion of the $\varepsilon$-$Ga_2O_3$ film. The latter one becomes essential as a blocking layer between the photocarriers in the upper illumination area and the defect traps near the bottom interface. As such, the micrometer-thick $\varepsilon$-$Ga_2O_3$ films achieve maximum photocarrier collection, and obtain simultaneous optimization of the responsivity and the response time. This study offers a unique perspective for engineering the defect profile in UWBG semiconductors with large minority carrier diffusivity for developing high-performance MSM DUV photodetectors.

## 4. METHODS

*Thin film growth*: The Sn-doped ε-$Ga_2O_3$ films with a 1% molar doping ratio were deposited on c-plane sapphire using pulsed laser deposition at a growth temperature of 750-800 °C and a dynamic oxygen pressure of 20 mTorr $O_2$. A 248 nm excimer laser source was used with a laser fluence of 1-2 J/cm$^2$ and a repetition rate of 2-5 Hz. After deposition, all the films were cooled down at 20 mTorr $O_2$ with a cooling rate of 20 °C/min. The film thickness was controlled by adjusting the laser pulse number.

*Structural, chemical and optical characterization:* The thin film phase, crystallinity and microstructure were investigated with high-resolution X-ray diffraction (XRD, Bruker D8 Advance) and transmission electron microscope (TEM, ThermoFisher Talos F200X) equipped with a Super-X energy-dispersive X-ray (EDX) analyzer. The film surface morphology was investigated using atomic force microscopy (AFM, Bruker Dimension Icon). The film thickness was mainly checked using the cross-sectional specimens by scanning electron microscopy (SEM, HITACHI Regulus-8230). The element oxidation state was characterized using X-ray photoemission spectroscopy (XPS, Kratos AXIS SUPRA) using Mg $K\alpha$ (h$v$ = 1253.6 eV) as the excitation source. For XPS depth profile measurement, an $Ar^+$ ion sputtering source operated at a 4 kV voltage was used with an interval sputtering time of 300 seconds. The beam scan size is 4×4 mm$^2$ and the incident angle is 45 degree. Optical transmission spectra were collected with a UV-Vis/NIR Spectrophotometer (PerkinElmer, Lambda 950).

*Photodetector fabrication and performance measurement:* Photodetector devices with a metal-semiconductor-metal (MSM) structure are fabricated using e-beam evaporation by depositing a Au (50 nm)/Ti (20 nm) bilayer as the interdigital electrodes on top of the $Ga_2O_3$ films. The electrode spacing and the effective electrode area were controlled using a direct-write optical lithography machine (MicroWriter ML3) and a lift-off process. The photodetector performance was evaluated using a Keithley 4200 parameter analyzer connected with a probe station at ambient conditions. The incident light with variable intensities was calibrated with a standard UV-enhanced Si photodiode sensor. For spectral response measurement, monochromatic illumination was obtained from a 300 W Xenon lamp equipped with a monochromator. The calculation methods for the photodetectors metrics, including responsitvity, detectvity and response time, are described in references.[4,5]

ASSOCIATED CONTENT

**Supporting Information**

The Supporting Information is available free of charge on the ACS Publications website.

Film thickness as a function of the laser pulse number for as-deposited ε-$Ga_2O_3$ films on sapphire. Tauc plot of the 2kp-thick ε-$Ga_2O_3$ film. Current-voltage (*I-V*) characteristics of the ε-$Ga_2O_3$ MSM photodetectors with different film thicknesses. . Dark *I-V* characteristics of the 20kp-thick ε-$Ga_2O_3$ MSM photodetectors with a $t_{space}$ of 1μm, 2μm and 4μm. High-resolution XPS depth profile analysis of the Ga 2p and Sn 3d core levels for 2kp-thick and 20kp-thick ε-$Ga_2O_3$ films. TEM characterization results of the ε-$Ga_2O_3$ film from the bottom region to the top region showing the evolution of misfit dislocations and domain boundaries through the film thickness. A supporting table of performance comparison of reported MSM photodetectors based on planar $Ga_2O_3$ films with different $Ga_2O_3$ phases

AUTHOR INFORMATION

**Corresponding Author**

*E-mail (Wenrui Zhang): zhangwenrui@nimte.ac.cn

*E-mail (Jichun Ye): jichun.ye@nimte.ac.cn

**Notes**

The authors declare no conflict of interest.

ACKNOWLEDGMENTS

This research was supported by Zhejiang Provincial Natural Science Foundation under Grant No. LZ21F040001, the Pioneer Hundred Talents Program of Chinese Academy of Sciences, Ningbo Yongjiang Talent Introduction Programme and Ningbo Key Scientific and Technological Project (Grant No. 2021Z083). Li Ji acknowledges the young scientist project of MOE innovation platform and State Key Laboratory of ASIC & System (2021MS004).

# Figures and captions

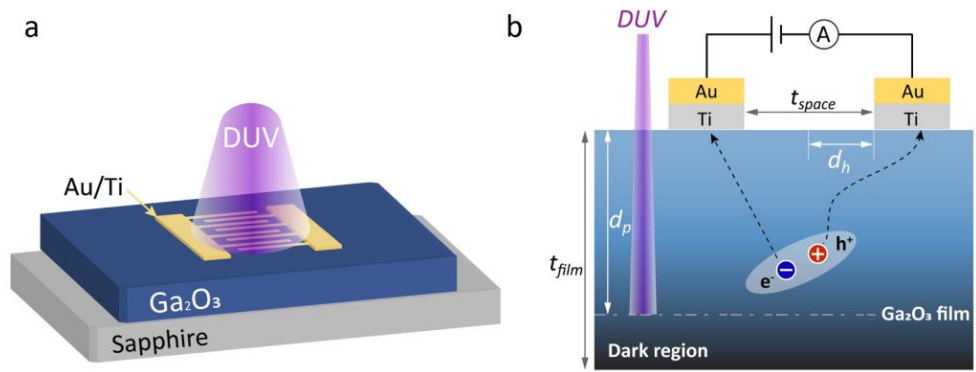

**Scheme 1.** (a) Schematic of a MSM photodetector for DUV light detection. (b) Schematic of a photodetection incident occurring in a MSM photodetector, which describes the key geometric parameters (film thickness $t_{film}$ and electrode spacing $t_{space}$) along with the material characteristics (effective light absorption length $d_p$ and the minority carrier diffusion length $d_h$).

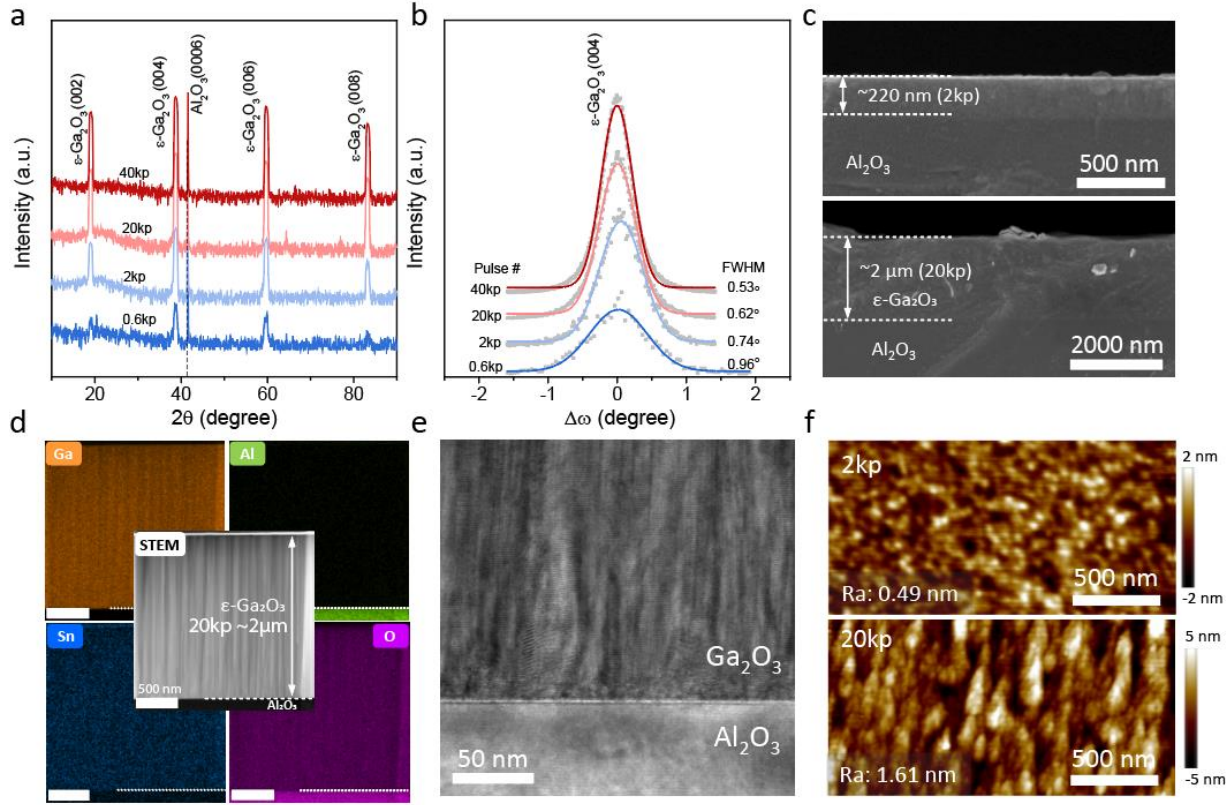

**Figure 1.** (a) θ-2θ XRD full scans and (b) rocking curve profiles of Sn-doped ε-Ga₂O₃ films with different film thicknesses grown on c-plane sapphire. Gaussian fitting described by the line plot is used to determine the FWHM value of the rocking curve data described by scattered points. (c) Cross-sectional SEM images of ε-Ga₂O₃ films with pulse numbers of 2kp and 20kp for the thickness measurement. (d) Cross-sectional HAADF-STEM image of the 20kp ε-Ga₂O₃ film and corresponding EDX maps for Ga, Al, Sn, O elements in the same region. (e) High-resolution TEM image of the ε-Ga₂O₃ film near the film-substrate interface showing a homogenous film lattice and an abrupt interface. (f) AFM images of the 2kp-thick and 20kp-thick films shows a smooth surface morphology.

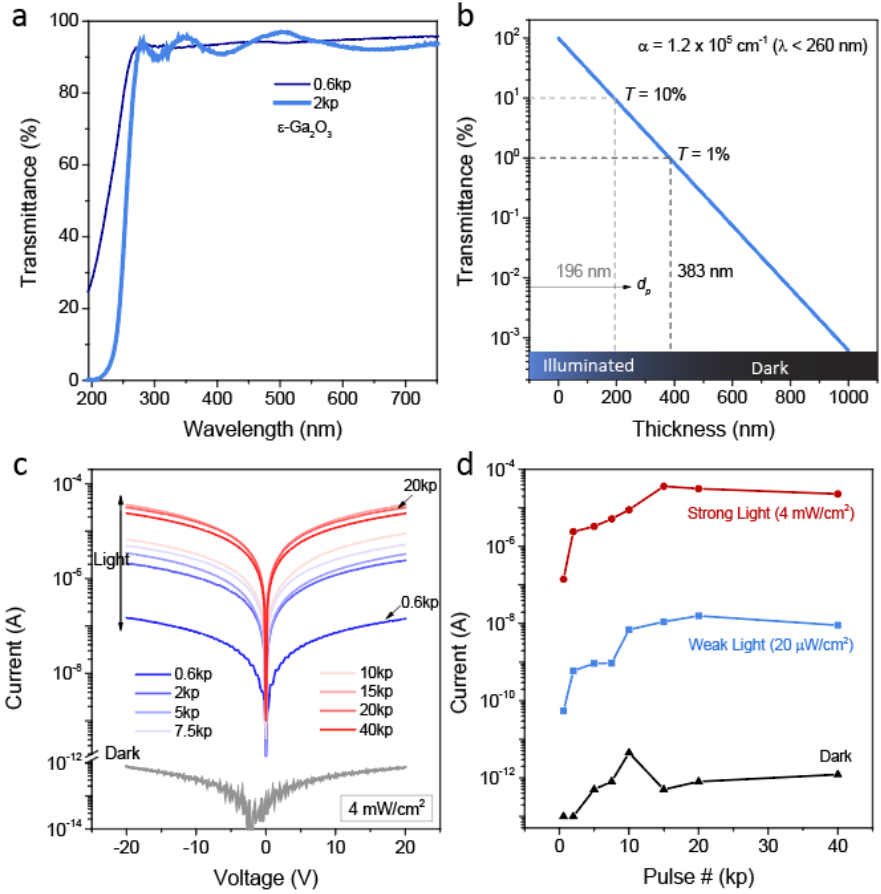

**Figure 2.** (a) Transmission spectra of the 0.6kp-thick and 2kp-thick ε-Ga₂O₃ films. (b) Thin film transmission as a function of the ε-Ga₂O₃ film thickness for the incident light with λ < 260 nm. (c) Current-voltage characteristics of the ε-Ga₂O₃ MSM photodetectors with different film thicknesses under 254 nm illumination with a light intensity of 4 mW/cm². The dark current of the 20kp-thick device is plotted as a comparison. (d) Photocurrent as a function of the pulse number at strong and weak light conditions, with corresponding dark current for comparison.

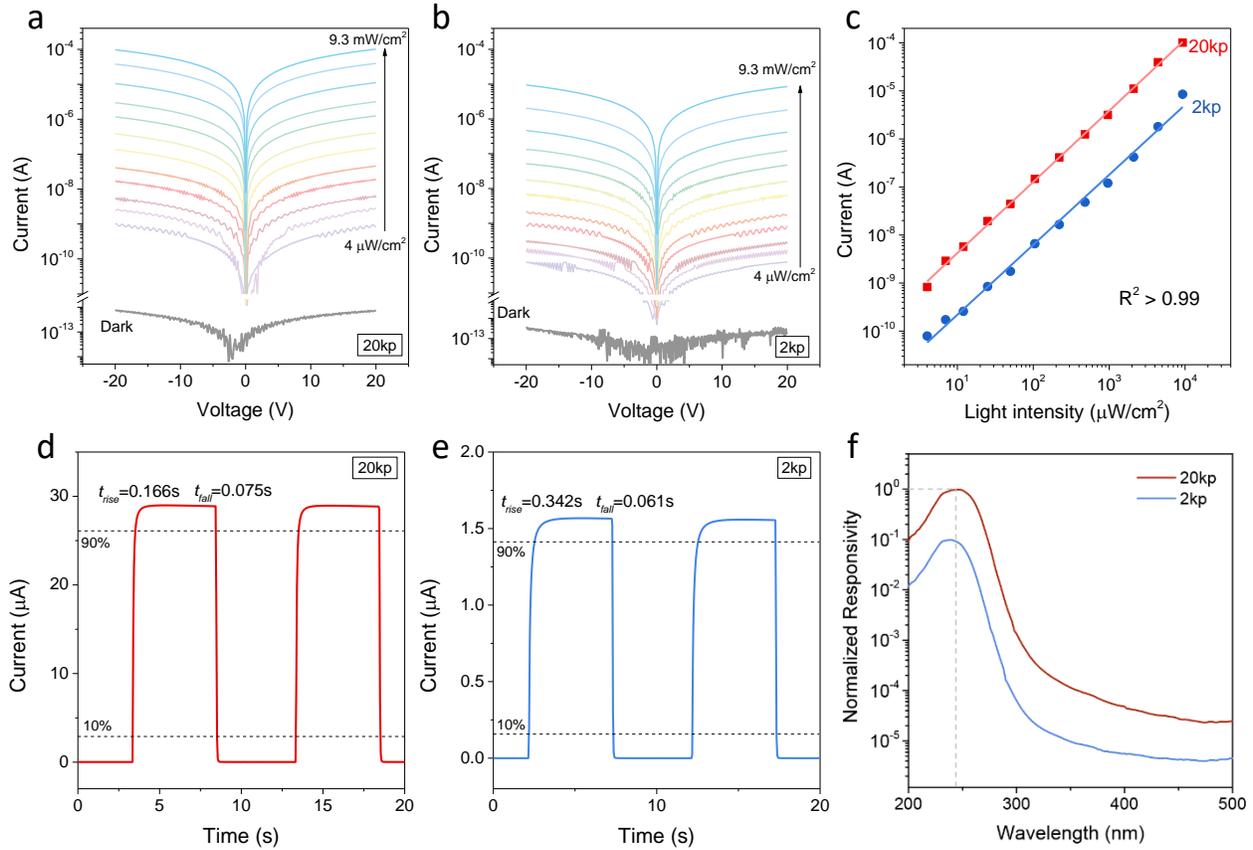

**Figure 3.** Current-voltage characteristics of (a) 20kp-thick and (b) 2kp-thick ε-Ga$_2$O$_3$ MSM photodetectors under different light intensities with corresponding dark current for comparison. (c) Photocurrent under a 20 V bias as a function of light intensity for 20kp-thick and 2kp-thick ε-Ga$_2$O$_3$ photodetectors. Current-time characteristics of (d) 20kp-thick and (e) 2kp-thick ε-Ga$_2$O$_3$ photodetectors under ON/OFF modulated illumination measured at 20 V for the dynamic response and reproductivity evaluation. The monochromatic illumination source has a wavelength of 254 nm and an intensity of 4 mW/cm$^2$. (f) Wavelength-dependence responsivity at 20 V for 20kp-thick and 2kp-thick ε-Ga$_2$O$_3$ photodetectors. The responsivity values are normalized based on the peak responsivity of the 20kp-thick ε-Ga$_2$O$_3$ photodetector.

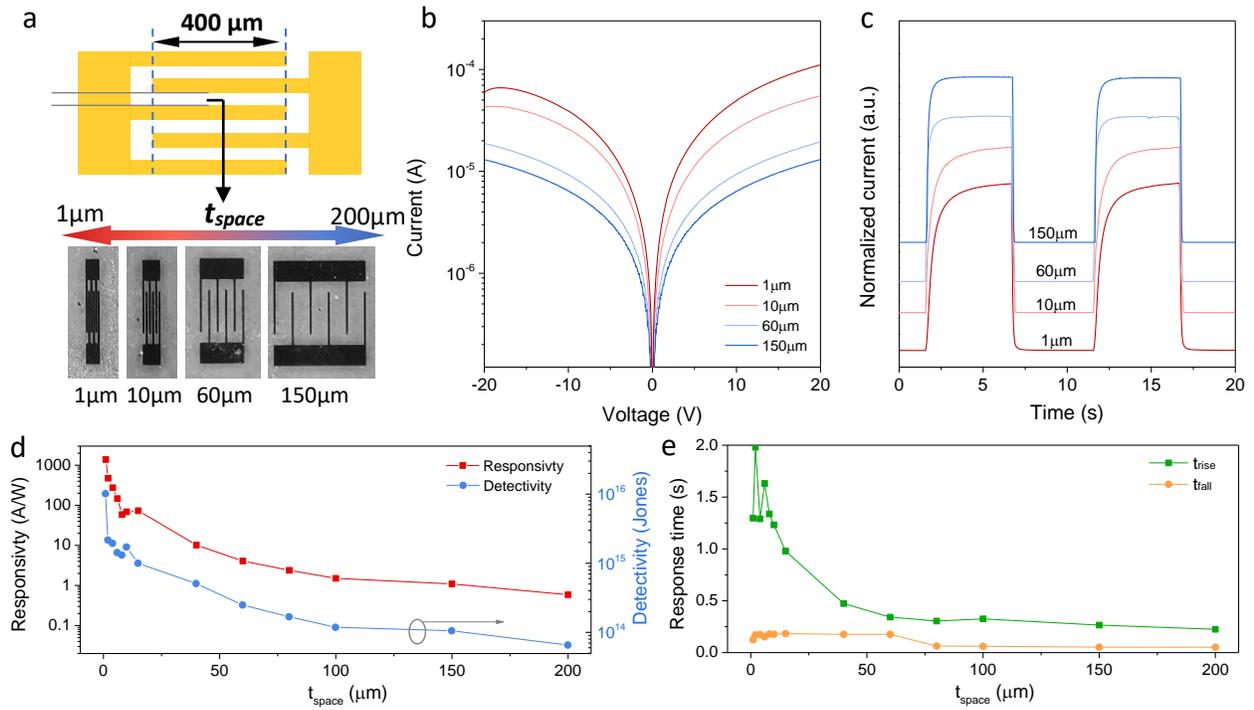

**Figure 4.** (a) Schematic of an interdigitated electrode design (upper) and corresponding optical images with variable $t_{space}$ (lower). (b) Current-voltage and (c) current-time characteristics of 20kp-thick $\varepsilon$-Ga$_2$O$_3$ MSM photodetectors with four different $t_{space}$ of 1 μm, 10 μm, 60 μm and 150 μm measured with a light intensity of 4 mW/cm$^2$. (d) Responsivity (left axis) and detectivity (right axis) and (e) response time as a function of $t_{space}$ from 1 μm to 200μm for the 20kp-thick $\varepsilon$-Ga$_2$O$_3$ MSM photodetector.

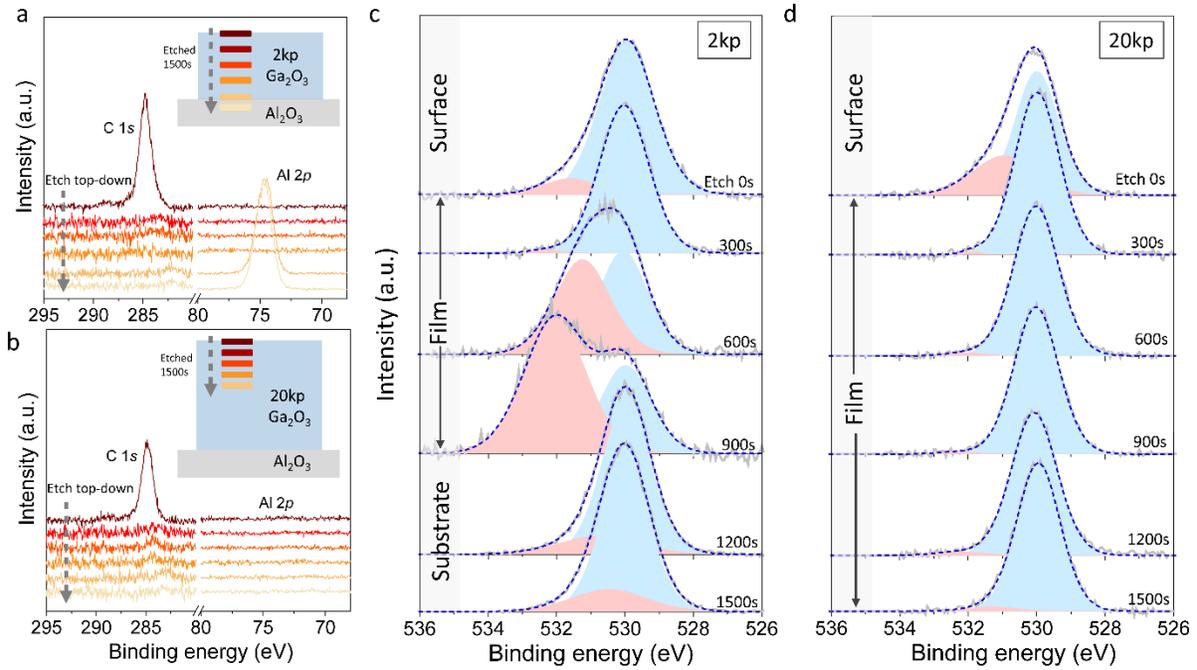

**Figure 5.** High-resolution C *1s* and Al *2p* XPS spectra collected at different etched depths for (a) 2kp- and (b) 20kp-thick ε-Ga$_2$O$_3$ films. The sputtering ion etching and data collection interval is 300 s, and the total etching time is 1500 s for both films. The depth profile analysis of the O *1s* core level for (c) 2kp- and (d) 20kp-thick ε-Ga$_2$O$_3$ films using the same etching depth, which respectively covers the entire film region for the 2kp film and only the top film region for the 20kp film.

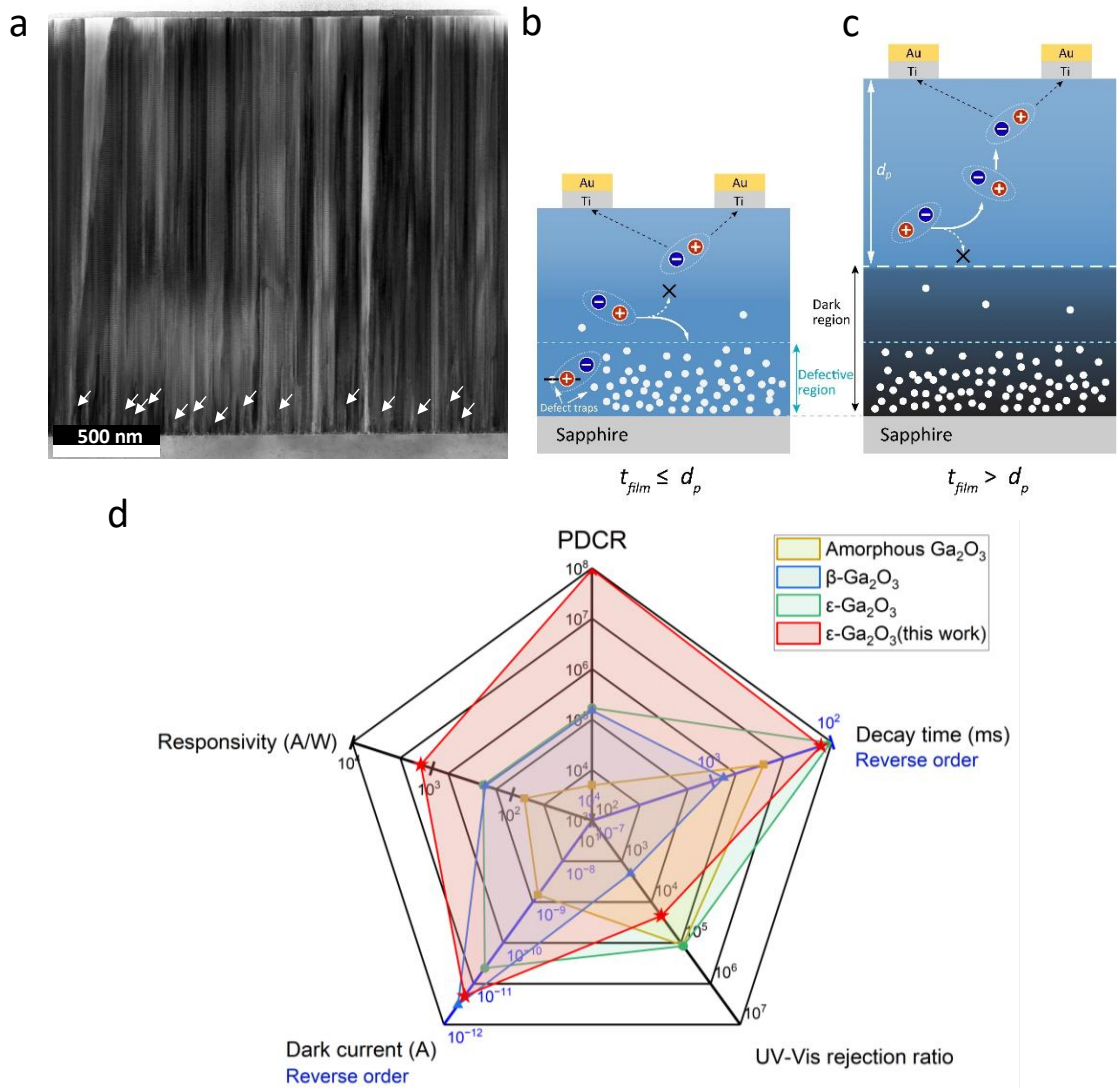

**Figure 6.** (a) Cross-sectional bright-field STEM image of a 20kp-thick ε-Ga$_2$O$_3$ film showing the formation of a defective region featured by a high density of misfit dislocations near the interface, followed by dislocation annihilation and domain broadening in the thicker region. Schematic of photocarrier transport for (b) $t_{film} \leq d_p$ and (c) $t_{film} > d_p$ in the ε-Ga$_2$O$_3$ MSM thin film photodetectors illustrating the critical role of the extra dark region in assisting carrier collection. (d) Radar plot summarizing critical device metrics of the ε-Ga$_2$O$_3$ MSM photodetector of this study compared with previous high-performance Ga$_2$O$_3$ MSM photodetectors based on different Ga$_2$O$_3$ phases.